\documentclass[12pt]{article}
\usepackage{amsmath}

\begin{document}
\begin{titlepage}

\title{A teleparallel type theory for massless spin 2 fields}

\author{J. W. Maluf$\,^{1}$ and S. C. Ulhoa$\,^{2}$ \\
Instituto de F\'{\i}sica, 
Universidade de Bras\'{\i}lia\\
70.919-970 Bras\'{\i}lia DF, Brazil\\}
\maketitle
\bigskip
\bigskip

\begin{abstract}
We present the Lagrangian and Hamiltonian formulations of a theory for spin 2
fields. The construction is developed in flat space-time. The construction in
curved space-time is conceptually straightforward, although it is not unique.
The theory is based on a symmetric tensor $S_{\mu\nu}$, contains two 
degrees of freedom of radiation, is motivated by the teleparallel formulation
of general relativity, and displays a certain resemblance with Maxwell's 
theory for the electromagnetic field.
\end{abstract}
\thispagestyle{empty}
\bigskip
\vfill
\begin{footnotesize}
\end{footnotesize}

\bigskip
{\footnotesize
\noindent {(1)} jwmaluf@gmail.com, wadih@unb.br\par
\noindent {(2)} sc.ulhoa@gmail.com}
\end{titlepage}

\newpage
\section{Introduction}
It is generally accepted that a theory for spin 2 fields is based on second
order symmetric tensors. However, there is no consensus as to what would be
the theory for these symmetric tensors. One standard approach for obtaining
field equations for massless spin 2 fields is the linearisation of Einstein's
equations. A second standard approach, although not completely independent 
from the former, is given by the Fierz-Pauli theory \cite{Fierz,Novello}.
In the case of massive spin 2 fields, there are also several investigations
in the literature (see, for instance, Refs. \cite{Watanabe,Buchbinder,deRham}
and references therein). Theories with slightly massive gravitons have 
attracted attention recently, as they could be relevant for the 
cosmological constant problem. Unlike the case of spin 1 fields described by 
Maxwell's theory, the interaction of spin 2 fields with other fields is not a
trivial issue, as it is not entirely clear what is the source (or sources)
of the massless spin 2 fields. Electric charges and currents are sources of 
the electromagnetic field, but a similar physical description is not known
to exist for spin 2 fields. As we will discuss at the end of this article, 
the gravitational field might be one source for spin 2 fields.

In this article we present a new theory for the dynamics of spin 2 fields.
The theory follows from the analysis developed in Ref. \cite{Maluf1}, and
is based on a Lagrangian density similar to the one adopted in the 
teleparallel equivalent of general relativity (TEGR) \cite{Maluf2,Maluf3},
the difference being that instead of using tetrad fields, as is normally
done in the TEGR, now one redefines the Lagrangian density in terms of a
symmetric tensor $S_{\mu\nu}$. By adopting a suitable gauge condition, the
symmetric tensor obeys the wave equation for massless fields. Although the
structure of the theory in consideration here is very much different from
the theory for spin 1 fields, it shares similarities with Maxwell's theory
for the electromagnetic field.

In section 2 we present the Lagrangian formulation of the theory, although
most of the content of this section was already presented in 
Ref. \cite{Maluf1}. In section 3, the Hamiltonian formulation of the theory
is developed. At the end of this section we show that the theory describes
two degrees of freedom of radiation, in similarity to the degrees of freedom
that arise in the linearisation of Einstein's equations, and that yield 
linearised gravitational waves. In section 4 we calculate the energy density
of the spin 2 fields, assuming the same gauge that is used in the 
linearisation of gravitational waves. In the final section we discuss the 
interaction of the spin 2 fields with the gravitational field. We show that
there is a natural minimal coupling with the gravitational field, but the
coupling seems to be different if the background space-time geometry is 
either the Weitzenb\"ock or Riemannian space.

\section{The Lagrangian formulation}

The geometrical framework of the theory considered in this article has already
been presented in Ref. \cite{Maluf1}. The motivation is to look for an
extension of the standard teleparallel equivalent of general 
relativity (TEGR) by considering
the distorted torsion tensor, a quantity that was 
introduced by Okubo \cite{Okubo3}. The distorted torsion tensor is defined in
terms of the tetrad fields and of a tensor $S^\lambda_\mu$ which, in turn, 
defines the Nijenhuis tensor \cite{NJ,Nakahara}. It was observed that in flat
space-time limit of the models considered in Ref. \cite{Maluf1}, 
the symmetric form of this tensor obeys the usual wave equation,
a feature that also holds in curved space-time. 

In the present analysis, the
focus is only on the tensor $S_{\lambda\mu}=S_{\mu\lambda}$, in order to 
investigate the essential features of the theory. For this purpose, 
we will treat the theory just like we address Maxwell's theory in the 
standard way. The theory considered below is formulated in flat space-time,
and is invariant under Lorentz transformations. The introduction of a flat 
space-time metric tensor, or flat space-time tetrad fields, to build a
theory invariant under general coordinate transformations, will be carried 
out in the future, only after the relevant aspects of the theory are
displayed and understood.

We will use, as much as possible, 
the same notation of Ref. \cite{Maluf1}. The flat space-time
metric tensor is $\eta_{\mu\nu}=(-1,+1,+1,+1)$, and 
$S_{\lambda\mu}=\eta_{\lambda\rho}\,S^\rho_\mu$, 
$S^{\rho\sigma}=S_\mu^\rho\,\eta^{\mu\sigma}$. We define the tensor

\begin{equation}
{\cal T}_{\lambda\mu\nu}=
\partial_\mu S_{\lambda\nu}-\partial_\nu S_{\lambda\mu}
=-{\cal T}_{\lambda\nu\mu}.
\label{1}
\end{equation}
The Lagrangian density for the theory of spin 2 fields is defined by

\begin{equation}
L=-\frac{1}{2} k\biggl(
{1\over 4} {\cal T}^{\lambda\mu\nu}{\cal T}_{\lambda\mu\nu}+
{1\over 2} {\cal T}^{\lambda\mu\nu}{\cal T}_{\mu\lambda\nu}-
{\cal T}^\lambda {\cal T}_\lambda\biggr)\,,
\label{2}
\end{equation}
where ${\cal T}_\mu={\cal T}^\lambda\,_{\lambda\mu}$, and $k=c^3/(16\pi G)$.
The Lagrangian density may be rewritten as 

\begin{equation}
L=-{1\over 2}k\Sigma^{\lambda\mu\nu}{\cal T}_{\lambda\mu\nu}\,,
\label{3}
\end{equation}
where

\begin{equation}
\Sigma^{\lambda\mu\nu} = {1\over 4}\left({\cal T}^{\lambda\mu\nu} + 
{\cal T}^{\mu\lambda\nu} - {\cal T}^{\nu\lambda\mu}\right) 
+ {1\over 2}\left(\eta^{\lambda\nu}{\cal T}^{\mu} - 
\eta^{\lambda\mu}{\cal T}^{\nu}\right)\,.
\label{4}
\end{equation}
Equations (\ref{2},\ref{3},\ref{4}) are similar to the quantities that arise
in the TEGR \cite{Maluf2,Maluf3}. In particular, Eq. (\ref{4}) was first 
established in Ref. \cite{Maluf2}. The field equations that are obtained from
the variation of the action integral are given by

\begin{equation}
k(\partial_\mu \Sigma^{\lambda\mu\nu}+\partial_\mu \Sigma^{\nu\mu\lambda})=0\,.
\label{5}
\end{equation}
The equation above is symmetric in the indices $(\lambda\nu)$, However, it can
be verified by explicit calculations that the quantity 
$\partial_\mu \Sigma^{\lambda\mu\nu}$ alone is symmetric in the indices 
$(\lambda\nu)$, if $\Sigma^{\lambda\mu\nu}$ is written in terms of 
Eq. (\ref{1}). For this reason, this equation can be simplified as 

\begin{equation}
2k\,\partial_\mu \Sigma^{\lambda\mu\nu}=0\,.
\label{6}
\end{equation}

We consider now a hypothetical interaction of the field $S_{\lambda\nu}$
with matter fields. Let us assume that the Lagrangian density for the matter
fields $\psi_M$  depends also on $S_{\lambda\nu}$, i.e., 
$L_M=L_M(\psi_M,S_{\lambda\nu})$. We define the energy-momentum tensor
$T^{\lambda\nu}$ according to

\begin{equation}
{\delta \over {\delta S_{\lambda\nu}}}\biggl( {1\over c} L_M\biggr)\equiv
{1\over c}T^{\lambda\nu}\,.
\label{7}
\end{equation}
where $c$ is the speed of light. We see no other interpretation for the 
tensor $T^{\lambda\nu}$ other than an energy-momentum type tensor. If 
$S^{\lambda\nu}$ is of gravitational nature, on equal footing with the tetrad 
fields (as discussed in the first model of Ref. \cite{Maluf1}), 
then it interacts with the matter fields as the gravitational field does,
possibly via the minimal coupling.
\footnote{As discussed in Ref. \cite{Maluf1}, 
possible solutions for the tensor $S_{\lambda\mu}$ are given in the form
$S_{\lambda \mu} = \eta_{\lambda \mu} +\texttt{wave solution}$ (see Eq.
(\ref{14}) below), and the 
projection of this tensor on a tetrad basis yields 
$S_{ab}=e_a\,^\lambda e_b\,^\mu S_{\lambda \mu}$, where $(a,b)$ are Lorentz
indices. This latter tensor may be interpreted as an extension 
of the flat Minkowski (tangent space) metric tensor 
$\eta_{ab}=(-1,+1,+1,+1)$, which includes oscillations (fluctuations) of the
background geometry. In this sense, the tensor $S_{\lambda\mu}$
might be of gravitational nature}
However, it is possible (and in fact, 
very likely) that the spin 2 field $S_{\lambda\mu}$ interacts only with
the gravitational field, according to the discussion in the last section
of the article. Thus, for the time being, we assume the validity of Eq. 
(\ref{7}). Then, in the presence of matter fields, the 
field equations for the spin 2 fields are assumed to be 

\begin{equation}
\partial_\mu \Sigma^{\lambda\mu\nu}=-{1\over {2kc}}T^{\lambda\nu}\,.
\label{8}
\end{equation}
Denoting $T$ as the trace of the tensor $T^{\lambda\nu}$, it follows from the 
equation above that 

\begin{equation}
\eta_{\lambda\nu}\,\partial_\mu \Sigma^{\lambda\mu\nu}=-{1\over {2kc}}T\,.
\label{9}
\end{equation}
With the help of this equation, and after some rearrangements, the field 
equation (\ref{8}) is simplified as 

\begin{equation}
\partial_\mu\partial^\mu S^{\lambda\nu}-
\partial_\mu\partial^\lambda S^{\mu\nu}
-\partial_\mu\partial^\nu S^{\mu\lambda}+
\partial^\lambda\partial^\nu S^\mu\,_\mu=
-{1\over {kc}}(T^{\lambda\nu}-{1\over 2}\eta^{\lambda\nu}T)\,.
\label{10}
\end{equation}
The left hand side of (\ref{10}) is 
similar to the linearised Ricci tensor for $h_{ab}$, which is considered
in the investigation of linearised gravitational waves, where
$g_{ab} \simeq\eta_{ab}+h_{ab}$ (see eq. (20.10) of 
Ref. \cite{Ray}; here we are adopting the notation of the latter reference,
where $(a,b,...)$ are
space-time indices). Note, however, that  Eq. (\ref{10}) is exact, since
it does not follow from any linearisation procedure.

It is straightforward do verify that the
field equations (\ref{10}) are invariant under the gauge transformation

\begin{equation}
S^{\lambda\mu}\rightarrow \tilde{S}^{\lambda\mu}=S^{\lambda\mu}+
\partial^\lambda V^\mu+\partial^\mu V^\lambda\,,
\label{11}
\end{equation}
where $V^\lambda(x)$ is an arbitrary vector field. We may use this gauge 
freedom to require the tensor $S^{\lambda\mu}$ to satisfy the condition

\begin{equation}
\partial_\rho S^{\lambda\rho}-{1\over 2} \partial^\lambda S^\rho\,_\rho=0\,.
\label{12}
\end{equation}
The condition above is known as the  de Donder, or Einstein, or Hilbert, 
or Fock gauge \cite{Ray}. By requiring $S^{\lambda\mu}$ to satisfy Eq. 
(\ref{12}), we arrive at (dropping the tilde)

\begin{equation}
\partial_\mu\partial^\mu S^{\lambda\nu}=
-{1\over {kc}}(T^{\lambda\nu}-{1\over 2}\eta^{\lambda\nu}T)\,.
\label{13}
\end{equation}
Thus, in empty space-time we obtain the wave equation

\begin{equation}
\partial_\mu \partial^\mu S^{\lambda\nu}=0\,.
\label{14}
\end{equation}

In view of the anti-symmetry 
$\Sigma^{\lambda\mu\nu}=-\Sigma^{\lambda\nu\mu}$, we have the conservation law

\begin{equation}
\partial_\mu T^{\lambda\mu}=0\,,
\label{15}
\end{equation}
that follows from Eq. (\ref{8}).
Finally, note that the tensor ${\cal T}_{\lambda\mu\nu}$ satisfies the cyclic
relation

\begin{equation}
{\cal T}_{\lambda\mu\nu}+{\cal T}_{\mu\nu\lambda}+
{\cal T}_{\nu\lambda\mu}=0\,,
\label{16}
\end{equation}
a property that the third rank tensors are required do satisfy in the
framework of the Fierz-Pauli theory \cite{Novello}.

\section{The Hamiltonian formulation}

We will develop this section in two parts. First, we carry out the Legendre 
transform, obtain the primary Hamiltonian and the primary constraints. In
the second part we establish the Poisson brackets, analyse the constraint
structure and write the time evolution equations for the field variables in 
the phase space of the theory. We show the equivalence of these equations 
with the Lagrangian field equations, and finally determine the number of the
dynamical degrees of freedom of the theory.

\subsection{The Legendre transform}

In order to carry out the Legendre transform of the Lagrangian density 
(\ref{3}), we find it more suitable to rewrite the latter in 
first order differential form. For this purpose, we introduce the new set of
fields $\phi_{\lambda\mu\nu}$, that eventually will be eliminated by the
field equations. Thus, we establish 
$L=L(S_{\lambda\mu}, \phi_{\lambda\mu\nu})$.

We define the tensor $\Lambda^{\lambda\mu\nu}$ in similarity to Eq. (\ref{4}),

\begin{equation}
\Lambda^{\lambda\mu\nu}={1\over 4}(\phi^{\lambda\mu\nu}+\phi^{\mu\lambda\nu}
-\phi^{\nu\lambda\mu})+{1\over 2}(\eta^{\lambda\nu}\phi^\mu-
\eta^{\lambda\mu}\phi^\nu)\,,
\label{17}
\end{equation}
where $\phi^\mu=\phi^\lambda\,_\lambda\,^\mu$. Thus, the Lagrangian density 
in first order differential formulation is given by

\begin{equation}
L={1\over 2}k\,\Lambda^{\lambda\mu\nu}(\phi_{\lambda\mu\nu}-
2{\cal T}_{\lambda\mu\nu})\,.
\label{18}
\end{equation}
The variation of $L$ with respect to $\phi_{\lambda\mu\nu}$ yields

\begin{equation}
\delta L=k\biggl[ (\delta \Lambda^{\lambda\mu\nu})(\phi_{\lambda\mu\nu}-
{\cal T}_{\lambda\mu\nu}) \biggr]=0\,,
\label{19}
\end{equation}
from what we conclude that $\phi_{\lambda\mu\nu}={\cal T}_{\lambda\mu\nu}$.
This equation splits into two parts,

\begin{eqnarray}
\phi_{\lambda 0 j}&=&{\cal T}_{\lambda 0 j} \label{20}\,, \\
\phi_{\lambda i j}&=&{\cal T}_{\lambda i j} \label{21}\,.
\end{eqnarray}
The Latin indices $i,j,k ....$ represent space indices and run from 1 to 3.
In the 3+1 decomposition of (\ref{18}), we will use Eq. (\ref{21}) to 
eliminate $\phi_{\lambda i j}$ from the very beginning, but the ``velocity"
fields $\phi_{\lambda 0 j}$ will be eliminated only at the end of the 
Legendre transform, after inverting the velocities in terms of the momenta.

As usual, the dot indicates time derivative, i.e., 
$\dot{S}_{\lambda\mu}=\partial_0 S_{\lambda\mu}$. In Eq. (\ref{18}), there is no
time derivative of the field quantity $S_{00}$. The canonical momenta are
obtained from the Lagrangian density through the term
$(-2k\Lambda^{\lambda 0 j}){\cal T}_{\lambda 0 j}$, i.e., 

\begin{eqnarray}
(-2k\Lambda^{\lambda 0 j})\dot{S}_{\lambda j}&=&
(-2k\Lambda^{ 0 0 j})\dot{S}_{0 j}+
\lbrack -k(\Lambda^{i 0 j}+\Lambda^{j 0 i})\rbrack \dot{S}_{ij} \\ \nonumber
&\equiv& \Pi^{0j}\dot{S}_{0j} + \Pi^{ij} \dot{S}_{ij}\,. \label{22}
\end{eqnarray}
Thus, we define the momenta canonically conjugated to $S_{0j}$ and $S_{ij}$ by

\begin{equation}
\Pi^{0j}=-2k \Lambda^{00j}\,,
\label{23}
\end{equation}

\begin{equation}
\Pi^{ij}=-k(\Lambda^{i0j}+\Lambda^{j0i})\,,
\label{24}
\end{equation}
respectively.
The right hand side of Eq. (\ref{23}) does not depend on time derivatives. 
It is equal to $-k{\cal T}^k\,_k\,^j$. As a consequence, we have the
following primary constraints,

\begin{eqnarray}
\Pi^{00}&=&0, \label{25} \\
\Pi^{0j}+k{\cal T}^k\,_k\,^j&=&0\,. \label{26} 
\end{eqnarray}

The algebra that leads to the expression of the ``velocities" in terms of
the momenta is not intricate. By developing Eq. (\ref{24}) we find

\begin{equation}
\Pi^{ij}=-{k \over 2} (\phi^{i0j}+\phi^{j0i})
-k\eta^{ij}(\eta^{kl}\phi_{k0l})\,.
\label{27}
\end{equation}
Denoting $\Pi=\Pi^i\,_i$, we obtain

\begin{equation}
\eta_{ij}\phi^{i0j}={1\over{2k}}\Pi\,,
\label{28}
\end{equation}
Thus, Eqs. (\ref{27}) and (\ref{28}) lead to

\begin{equation}
\phi^{i0i}+\phi^{j0i}=-{2\over k}(\Pi^{ij}-{1\over 2} \eta^{ij} \Pi)\,.
\label{29}
\end{equation}
From the equation above we obtain

\begin{equation}
\dot{S}_{ij}={1\over 2}(\partial_iS_{j0}+\partial_j S_{i0})+
{1\over k}(\Pi_{ij}-{1\over 2} \eta_{ij} \Pi)\,,
\label{30}
\end{equation}
and

\begin{equation}
\phi_{i0j}={1\over 2} {\cal T}_{0ij}+
{1\over k}(\Pi_{ij}-{1\over 2} \eta_{ij} \Pi)\,.
\label{31}
\end{equation}
With the help of the expressions above, it is not difficult to rewrite the
Lagrangian density in terms of the field variables in the phase space of the
theory. After a number of rearrangements and simplifications, we arrive at

\begin{eqnarray}
L&=& \Pi^{ij}\dot{S}_{ij} + \Pi^{0j}\dot{S}_{0j} -
{1\over {2k}}(\Pi^{ij}\Pi_{ij}-{1\over 2} \Pi^2)  \nonumber \\
&{}&-\Pi^{0j}\partial_j S_{00} -\Pi^{ij} \partial_j S_{i0} 
-{3\over 8}k {\cal T}^{0ij}{\cal T}_{0ij} 
-{1\over 2}k \Sigma^{kij}{\cal T}_{kij}\,.
\label{32}
\end{eqnarray}
Equation (\ref{32}) finally leads to the expression of the primary 
Hamiltonian,

\begin{eqnarray}
H_0&=& {1\over {2k}}(\Pi^{ij}\Pi_{ij}-{1\over 2} \Pi^2)+
\Pi^{ij} \partial_j S_{i0}+\Pi^{0j}\partial_j S_{00} \nonumber \\
&{}&+{{3k}\over 8} {\cal T}^{0ij}{\cal T}_{0ij}+
{k\over 2} \Sigma^{kij}{\cal T}_{kij}\,.
\label{33}
\end{eqnarray}
Both in (\ref{32}) and in (\ref{33}), the tensor $\Sigma^{kij}$ depends only
on ${\cal T}_{kij}$, i.e., it does not contain the ``$0$" index.

\subsection{Poisson brackets and time evolution equations}

The Poisson brackets between any two functions in the phase space of the 
theory is given by

\begin{equation}
\lbrace F, G\rbrace=\int d^3z\biggl(
{{\delta F}\over {\delta S_{\mu\nu}(z)}} 
{{\delta G}\over {\delta \Pi^{\mu\nu}(z)}}-
{{\delta F}\over {\delta \Pi^{\mu\nu}(z)}} 
{{\delta G}\over {\delta S_{\mu\nu}(z)}} \biggr)\,.
\label{34}
\end{equation}
With the help of the Poisson brackets, we will study the time evolution of
the primary constraints. Considering first Eq. (\ref{25}), we have

\begin{equation}
\dot{\Pi}^{00}(x)=\lbrace \Pi^{00}(x), \int d^3y H_0(y)\rbrace\,.
\label{35}
\end{equation}
After simple calculations, we obtain

\begin{equation}
\dot{\Pi}^{00}(x)=\partial_j \Pi^{0j}\simeq 0\,.
\label{36}
\end{equation}
Therefore, $\partial_j \Pi^{0j}=0$ is a secondary constraint of the theory.
The primary constraint (\ref{26}) also yields a secondary constraint. 
After long but also simple calculations, we obtain

\begin{eqnarray}
{d \over {dt}}(\Pi^{0i}(x)+k{\cal T}^k\,_k\,^i(x))&=&
\lbrace \Pi^{0i}(x)+k{\cal T}^k\,_k\,^i(x) , \int d^3y H_0(y)\rbrace 
\nonumber \\
&=&2\partial_j\Pi^{ij}(x)+k\partial_j{\cal T}^{0ji}(x) \nonumber \\
& \simeq & 0\,.
\label{37}
\end{eqnarray}

Therefore, we have the following set of constraints,

\begin{eqnarray}
A(x)&=& \Pi^{00}(x)=0\,,  \label{38}\\
B^i(x)&=& \Pi^{0i}(x)+k{\cal T}^k\,_k\,^i(x)\,, \label{39} \\
C(x) &=&\partial_j\Pi^{0j}(x)\,, \label{40} \\
D^j(x)&=& 2\partial_i\Pi^{ij}(x)+k\partial_i{\cal T}^{0ij}(x)\,, \label{41}
\end{eqnarray}
We recall that $A(x)$ and $B^i(x)$ are primary constraints, and $C(x)$ and 
$D^j(x)$ are secondary constraints. The time evolution of the constraint
$C(x)$ yields $\partial_i \partial_j \Pi^{ij}\simeq 0$, which is a 
consequence of the constraint (\ref{41}) (in view of the anti-symmetry 
${\cal T}^{0ij}=-{\cal T}^{0ji}$). Note also that by taking the divergence 
of the constraint $B^i$, and considering the constraint $C(x)$, we have
$\partial_i{\cal T}^k\,_k\,^i=0$. It can be verified by explicit calculations
that the resulting equation, 
$\partial_i\partial_k S^{ik}-\partial^k\partial_k S^i\,_i=0$,
is the $00$ component of the field equations (\ref{10}), assuming the 
vanishing of the energy-momentum tensor on the right hand side of (\ref{10}).
Finally, the time evolution of the constraint $D^j(x)$ is strongly equal to 
zero.

It can be shown that all possible Poisson brackets between the 
constraints (\ref{38}), (\ref{39}), (\ref{40}) and (\ref{41}) 
are strongly equal to zero.
In the evaluation of $\lbrace B^i(x), D^j(y) \rbrace$, special attention is
needed in the symmetrization of the variations required by Eq. (\ref{34})
(i.e., $\delta / \delta S_{ij}(z)=\delta / \delta S_{ji}(z)$).
Thus, all constraints of the theory are first class constraints.

In principle, one would have to add the first class constraints to the 
primary Hamiltonian density $H_0(x)$, each one multiplied by {\it a priori}
arbitrary Lagrange multipliers. The latter should be consistent with the 
field equations of the theory. However, it can be verified that the primary
Hamiltonian alone, constructed out of $H_0(x)$ only, already yields field 
equations that are strictly equivalent to the Lagrangian field equations.
First, it is very easy to show that Hamiltonian field equation

\begin{equation}
\dot{S}_{ij}=\lbrace S_{ij}(x), \int d^3y\,H_0(y)\rbrace\,,
\label{42}
\end{equation}
yields exactly Eq. (\ref{30}). As for the field equation for the momenta 
$\Pi^{ij}=\Pi^{ji}$, it is not difficult to conclude that

\begin{eqnarray}
\dot{\Pi}^{ij}&=&\lbrace \Pi^{ij}(x), \int d^3y\,H_0(y)\rbrace \nonumber \\
&=&k(\partial_m \Sigma^{imj}+\partial_m \Sigma^{jmi})\,.
\label{43}
\end{eqnarray}
Let us now identify $\phi_{i0j}={\cal T}_{i0j}$, which follows from 
Eq. (\ref{20}), and consider this identification in Eq. (\ref{24}). Then, 
it is straightforward to rewrite the time derivative of the momenta 
canonically conjugated to $S_{ij}$ as

\begin{equation}
\dot{\Pi}^{ij}=-k \partial_0(\Sigma^{i0j}+\Sigma^{j0i})\,.
\label{44}
\end{equation}
Combining the expression above with the right hand side of (\ref{43}), we find

\begin{equation}
k\partial_\mu (\Sigma^{i\mu j}+\Sigma^{j\mu i})=0\,,
\label{45}
\end{equation}
which is the $ij$ component of the field equation (\ref{8}), assuming the 
vanishing of the energy-momentum tensor in the latter. Considering the 
dependence of the tensor ${\cal T}_{\lambda\mu\nu}$ on $S_{\mu\nu}$, as in
Eq. (\ref{1}), then it is not difficult to show that each term
$\partial_\mu \Sigma^{i\mu j}$ and $\partial_\mu \Sigma^{j\mu i}$ are,
separately, symmetric in the indices $ij$. Making use of Eq. (\ref{1}), 
the field equation above is rewritten as 

\begin{eqnarray}
&{}&k(\partial_\mu \partial^\mu S^{ij}-\partial^j\partial_\mu S^{i\mu}-
\partial^i\partial_\mu S^{j\mu} \nonumber \\
&{}&+\eta^{ij}\partial_\mu \partial_\lambda S^{\lambda\mu}-
\eta^{ij}\partial_\mu \partial^\mu S^\lambda\,_\lambda+
\partial^i\partial^j S^\lambda\,_\lambda)=0\,.
\label{46}
\end{eqnarray}
This equation is invariant under the gauge transformation (\ref{11}), exactly
like Eq. (\ref{10}), again assuming $T^{\lambda\mu}=0$ in the latter.

The remaining Hamiltonian field equations are

\begin{eqnarray}
\dot{S}_{00}&=&\lbrace S_{ij}(x), \int d^3y\,H_0(y)\rbrace =0\,, 
\label{47} \\
\dot{S}_{0i}&=&\lbrace S_{0i}(x), \int d^3y\,H_0(y)\rbrace
=\partial_i S_{00}\,,
\label{48} \\
\dot{\Pi}^{0i}&=&\lbrace \Pi^{0i}(x), \int d^3y\,H_0(y)\rbrace =
\partial_j \Pi^{ij} +{{3k}\over 2}\partial_j {\cal T}^{0ji}\,.
\label {49}
\end{eqnarray}

We may now calculate the radiating degrees of freedom of the theory. In the
phase space of the theory we start with 10+10=20 degrees of freedom, and 
we have 8 first class constraints at each space-time event, as given by Eqs.
(\ref{38})-(\ref{41}). These constraints generate 
the expected symmetries and remove 8+8
degrees of freedom in the phase space. Thus, there remains 4 degrees of 
freedom in the phase space of the theory. In the configuration space, the
counting is the same. We start with 10 degrees of freedom, and observe
that the gauge symmetry (\ref{11}) removes 4 degrees of freedom, either from
Eq. (\ref{10}) or (\ref{46}). The equations 

\begin{eqnarray}
\ddot{S}_{00}&=&0\,, \label{50} \\
\ddot{S}_{0i}&=&0\,, \label{51} 
\end{eqnarray}
that follow from (\ref{47}) and (\ref{48}), also remove 4 degrees of freedom
(the Hamiltonian density may be rewritten such that the components 
$S_{00}$ and $S_{0i}$ arise as Lagrange multipliers, but the form of $H_0$ 
is more useful for our purposes),
and therefore we are left with 2 dynamical degrees of freedom in the
configuration space.

\section{The energy of the spin 2 fields}

The energy-momentum of the spin 2 fields described in this article is a 
relevant topic that deserves a careful analysis, which will be carried out
elsewhere. Here, we will just assume that the integral of the primary 
Hamiltonian density $H_0$ yields the energy of the fields.
In order to evaluate this energy, let us first recall that the left hand side
of Eq. (\ref{10}) is exactly equivalent to the linearised Ricci tensor for 
$h_{ab}$, where $g_{ab} \simeq\eta_{ab}+h_{ab}$ (in the notation of Ref.
\cite{Ray}), and which is considered in the investigation of linearised 
gravitational waves. Indeed, 
equation (20.10) of Ref. \cite{Ray} is constructed out of $h_{ab}$,
and is identical to the left hand side of (\ref{10}) (which is not 
linearised). Thus, we may adopt the gauge condition given by Eq. (\ref{12}),
which is also adopted in Ref. \cite{Ray}. In the latter reference, it is 
found the general form of the tensor $h_{ab}$ that satisfies a 
gauge condition similar to Eq. (\ref{12}). It is given by Eq. (20.48) of
Ref. \cite{Ray}, and reads

\begin{equation}
h_{ab}=
\begin{pmatrix}
h_{00} & -{1\over 2}(h_{00}+h_{11}) & h_{02} & h_{03} \\
-{1\over 2}(h_{00}+h_{11}) &  h_{11} & -h_{02} & -h_{03} \\
h_{02} & -h_{02} & h_{22} & h_{23} \\
h_{03} & -h_{03} & h_{23} & -h_{22} 
\end{pmatrix}\,.
\label{52}
\end{equation}
Making now the identification $S_{\lambda\mu}\leftrightarrow h_{ab}$, and
requiring initial conditions such that $S_{00}=S_{0i}=0$, we obtain

\begin{equation}
S_{\mu\nu}=
\begin{pmatrix}
0&0&0&0\\
0&0&0&0\\
0&0&S_{22}&S_{23}\\
0&0&S_{23}&-S_{22}\\
\end{pmatrix}\,.
\label{53}
\end{equation}
Both $S_{22}$ and $S_{23}$ are taken to be functions of $(t-x)$, assuming
that the fields propagate in the $x$ direction. The field equations for these
components are 
$\partial_\mu \partial^\mu\, S_{22}=\partial_\mu \partial^\mu\, S_{23}=0$.

Since $S_{\lambda\mu}$ reduce to only $S_{22}(t-x)$ and $S_{23}(t-x)$, we find

\begin{equation}
\Sigma^{iik}{\cal T}_{ijk}=2\left[\left(\partial_{x}S_{22}\right)^{2}
+\left(\partial_{x}S_{23}\right)^{2}\right]	\,,
\label{54}
\end{equation}
and

\begin{equation}
\Pi^{ij}\Pi_{ij}-\frac{1}{2}\Pi^{2}=
2k^{2}\left[\left(\partial_{0}S_{22}\right)^{2}+
\left(\partial_{0}S_{23}\right)^2\right]\,,
\label{55}
\end{equation}
which eventually yields

\begin{equation}
H_0=E=2k\left[\left(\partial_{0}S_{22}\right)^{2}+
\left(\partial_{0}S_{23}\right)^{2}\right]\,,
\label{56}
\end{equation}
where $2k=c^3/(8\pi G)$. The resulting expression is very simple, and should
represent the energy density of gravitons in a hypothetical quantum theory.

\section{Final comments}

We have analysed the Lagrangian and Hamiltonian formulations of
a theory for spin 2 fields. The structure of the theory is
motivated by the teleparallel equivalent of general relativity, and some 
general features of the theory remind Maxwell's theory for the 
electromagnetic field, as for instance the field equations (\ref{6}).
The conclusion is that the theory is consistent.
It is clear from the Hamiltonian analysis that the theory contains two
dynamical degrees of freedom in the configuration space.

One important aspect of the theory is the interaction of the spin 2 fields
with ordinary matter fields. This issue is not completely settled in this
article. However, we may envisage the interaction of the spin 2 fields with
the gravitational field. The natural coupling of the tensor field $S_{\mu\nu}$
with the gravitational field has already been addressed in Ref. \cite{Maluf1}.
We expect that the minimal coupling between $S_{\mu\nu}$ and either the 
tetrad fields or the metric tensor is given by equations (28) or (35) of
the latter reference, which read

\begin{equation}
\nabla_\mu(e\Sigma_\lambda\,^{\mu\nu})=
\partial_\mu(e\Sigma_\lambda\,^{\mu\nu})
-e\,\Gamma^\sigma_{\mu\lambda}\Sigma_\sigma\,^{\mu\nu}=0\,,
\label{57}
\end{equation}
where $e=\sqrt{-g}\;$, $g=\det(g_{\mu\nu})$, and which generalize Eq.
(\ref{6}).  In principle, the connection
$\Gamma^\sigma_{\mu\lambda}$ in the equation above may be either the 
Christoffel symbols, or the Weitzenb\"ock connection. This equation will be 
investigated in the weak field approximation of the gravitational field,
in which case the gravitational field acts as a source of the field 
equations (\ref{8}) or (\ref{10}). Therefore, time varying gravitational
fields generate the propagation of spin 2 fields.

A new aspect of the theory discussed in this article is the possible 
existence of static solutions of the field equations (\ref{10}), in 
similarity to the solutions of Gauss law in Maxwell's theory. By
investigating whether these solutions exist, and whether they are 
physical or not, we may have clues about the nature of the possible 
sources for the spin 2 fields.

\end{document}